\documentstyle[12pt]{article}
\begin{document}
\begin{titlepage}
\begin{flushright}
IC/2001/32\\
hep-th/0105093
\end{flushright}
\vspace{10 mm}

\begin{center}
{\Large The Cardy-Verlinde Formula and\\ 
Topological AdS-Schwarzschild Black Holes}

\vspace{5mm}

\end{center}

\vspace{5 mm}

\begin{center}
{\large Donam Youm\footnote{E-mail: youmd@ictp.trieste.it}}

\vspace{3mm}

ICTP, Strada Costiera 11, 34014 Trieste, Italy

\end{center}

\vspace{1cm}

\begin{center}
{\large Abstract}
\end{center}

\noindent

We consider the brane universe in the background of the topological 
AdS-Schwarzschild black holes.  The induced geometry of the brane is 
that of a flat or an open radiation dominated FRW-universe.  Just like 
the case of a closed radiation dominated FRW-universe, the temperature 
and entropy are simply expressed in terms of the Hubble parameter and 
its time derivative when the brane crosses the black hole horizon.  
We propose the modified Cardy-Verlinde formula which is valid for any 
values of the curvature parameter $k$ in the Friedmann equations.

\vspace{1cm}
\begin{flushleft}
May, 2001
\end{flushleft}
\end{titlepage}
\newpage

Verlinde made an interesting proposal \cite{ver1,ver2} that the Cardy 
formula \cite{car} for a two-dimensional conformal field theory (CFT) 
can be generalized to an arbitrary spacetime dimensions and such generalized 
formula, called the Cardy-Verlinde formula, is closely related to the 
Friedmann equation at the moment when the cosmological bounds on the 
thermodynamic quantities of the holographic dual theory are saturated.  
This result is later shown to hold for the holographic duals to various 
bulk backgrounds \cite{was,kps1,cai,bm,bim,kps2,youm,cai2}.  The quantum 
effects to the Cardy-Verlinde formula were studied in Refs. 
\cite{odi1,odi2,odi3}.  In this note, we consider the brane universe in the 
bulk background of the AdS-Schwarzschild black holes with the event horizon 
having zero and negative curvatures, the so-called topological black holes.  
(Cf. Some aspects of the brane cosmology in such bulk background were explored 
also in Refs. \cite{myu1,myu2}.)  The brane universes under consideration 
are therefore flat and open universes.  We propose the modified 
Cardy-Verlinde formula which holds for any values of the curvature parameter 
$k$ in the Friedmann equations.  We show that even for flat and open brane 
universes the proposed cosmological bounds on the temperature and entropy of 
the holographic dual theory are saturated and thereby simply expressed in 
terms of the Hubble parameter and its time derivative when the brane crosses 
the horizon of the AdS-Schwarzschild black hole.  

It is believed that black holes in asymptotically flat spacetime should 
have spherical horizon \cite{haw,fsw}.  However, when a spacetime has 
negative cosmological constant, a black hole can have non-spherical horizon.  
Four-dimensional black hole solutions whose horizon is an arbitrary genus 
Riemann surface were studied in Refs. \cite{man,blp,van,man1,absp,bri,kmv}.  
In Ref. \cite{bir}, such black hole solutions were generalized to arbitrary 
spacetime dimensions.  The solution has the following form:
\begin{eqnarray}
ds^2_{n+2}&=&-h(a)dt^2+{1\over{h(a)}}da^2+a^2\gamma_{ij}(x)dx^idx^j,
\cr
h(a)&=&k-{{w_{n+1}M}\over a^{n-1}}+{a^2\over L^2},\ \ \ \ \ \ 
\omega_{n+1}={{16\pi G_{n+2}}\over{n{\rm Vol}(M^n)}},
\label{adsbh}
\end{eqnarray}
where $\gamma_{ij}$ is the horizon metric for a constant curvature manifold 
$M^n$ with the volume ${\rm Vol}(M^n)=\int d^nx\sqrt{\gamma}$, $G_{n+2}$ is 
the $(n+2)$-dimensional Newton's constant, $M$ is the ADM mass of the black 
hole and $L$ is the curvature radius of the background AdS spacetime.  The 
horizon geometry of the black hole is elliptic, flat and hyperbolic for 
$k=1,0,-1$, respectively.  The Bekenstein-Hawking entropy and the Hawking 
temperature of the black hole are
\begin{equation}
S={{a^n_H{\rm Vol}(M^n)}\over{4G_{n+2}}},\ \ \ \ \ \ \ \ \ \ \ \ 
{\cal T}={{h^{\prime}(a_H)}\over{4\pi}}={{(n+1)a^2_H+(n-1)kL^2}\over
{4\pi L^2a_H}},
\label{tempent}
\end{equation}
where $a_H$ is the horizon, defined as the largest zero of $h(a)$, and 
the prime denotes the derivative w.r.t. $a$.  As for the $k=-1$ case, 
the requirement of positivity of temperature enforces an inequality on 
the value of $a_H$, namely that $a_H>L\sqrt{(n-1)/(n+1)}$.  

We consider an $n$-brane moving in the background of the above 
AdS-Schwarzschild black hole.  The metric on the brane is given by the 
induced metric
\begin{equation}
ds^2_{n+1}=-\left[h(a)-{1\over{h(a)}}\left({{da}\over{dt}}\right)^2
\right]dt^2+a^2\gamma_{ij}dx^idx^j.
\label{indmet}
\end{equation}
In terms of a new time coordinate $\tau$, called the cosmic time, satisfying
\begin{equation}
{1\over{h(a)}}\left({{da}\over{d\tau}}\right)^2-h(a)\left({{dt}\over
{d\tau}}\right)^2=-1,
\label{costim}
\end{equation}
the brane metric (\ref{indmet}) takes the standard Robertson-Walker form
\begin{equation}
ds^2_{n+1}=-d\tau^2+a^2(\tau)\gamma_{ij}dx^idx^j,
\label{rwmet}
\end{equation}
with the cosmic scale factor $a$.  The equation of motion for the brane 
action can be translated into \cite{ver2}
\begin{equation}
{{dt}\over{d\tau}}={{\sigma a}\over{h(a)}},
\label{eqbranact}
\end{equation}
where the parameter $\sigma$ is related to the brane tension.  From Eqs. 
(\ref{costim},\ref{eqbranact}), we obtain the following Friedmann equation 
for the radiation dominated brane universe:
\begin{equation}
H^2={{\omega_{n+1}M}\over a^{n+1}}-{k\over a^2}, 
\label{frdeq1}
\end{equation}
by fine-tuning the brane tension to $\sigma=1/L$ so that the cosmological 
constant term in the Friedmann equation vanishes.  Here, $H\equiv \dot{a}/a$ 
is the Hubble parameter, where the overdot denotes the derivative w.r.t. 
$\tau$.  Taking the $\tau$-derivative of Eq. (\ref{frdeq1}), we obtain the 
second Friedmann equation
\begin{equation}
\dot{H}=-{{n+1}\over 2}{{\omega_{n+1}M}\over{a^{n+1}}}+{k\over a^2}.
\label{frdeq2}
\end{equation}

According to the AdS/CFT correspondence, thermodynamic quantities of the 
CFT at high temperature can be identified with the corresponding 
thermodynamic quantities of the bulk AdS black hole \cite{wit}.   Since 
the standard GKPW prescription \cite{gkp,wit2} does not fix the overall 
scale of the boundary metric, we are free to re-scale the boundary metric 
to be of the following form:
\begin{equation}
ds^2_{CFT}=\lim_{a\to\infty}\left[{L^2\over a^2}ds^2_{n+2}\right]
=-dt^2+L^2\gamma_{ij}dx^idx^j.
\label{bndrmet}
\end{equation}
Since the CFT time is rescaled by the factor $L/a$ w.r.t. the AdS time, 
the energy $E$ and the temperature $T$ of the CFT are rescaled by the same 
factor w.r.t. the corresponding thermodynamic quantities of the AdS black 
hole:
\begin{equation}
E=M{L\over a},\ \ \ \ \ \ \ \ \ \ \ \ 
T={\cal T}{L\over a}={1\over{4\pi a}}\left[(n+1){a_H\over L}+(n-1){{kL}
\over a_H}\right],
\label{entempads}
\end{equation}
whereas the entropy $S$ of the CFT is given by the Bekenstein-Hawking 
entropy (\ref{tempent}) of the AdS black hole without re-scaling.  Note, 
in terms of the energy density $\rho=E/V$ and the pressure $p=\rho/n$ of 
the CFT within the volume $V=a^n{\rm Vol}(M^n)$, the Friedmann equations 
(\ref{frdeq1},\ref{frdeq2}) take the following standard forms:
\begin{equation}
H^2={{16\pi G}\over{n(n-1)}}\rho-{k\over a^2},
\label{frd1}
\end{equation}
\begin{equation}
\dot{H}=-{{8\pi G}\over{n-1}}(\rho+p)+{k\over a^2},
\label{frd2}
\end{equation}
where $G=(n-1)G_{n+2}/L$ is the Newton's constant on the brane.  From these 
Friedmann equations, we obtain the energy conservation equation $\dot{\rho}
+nH(\rho+p)=0$.  

The Friedmann equations (\ref{frd1},\ref{frd2}) can be respectively put into 
the following forms, resembling the formulas for the CFT:
\begin{equation}
S_H={{2\pi}\over n}a\sqrt{E_{BH}(2E-kE_{BH})},
\label{fredcft1}
\end{equation}
\begin{equation}
kE_{BH}=n(E+pV-T_HS_H),
\label{fredcft2}
\end{equation}
in terms of the Hubble entropy $S_H$ and the Bekenstein-Hawking energy 
$E_{BH}$, where
\begin{equation}
S_H\equiv (n-1){{HV}\over{4G}}, \ \ \ \ \ \ \ 
E_{BH}\equiv n(n-1){V\over{8\pi Ga^2}}, \ \ \ \ \ \ \ 
T_H\equiv -{\dot{H}\over{2\pi H}}.
\label{defs}
\end{equation}  
The first Friedmann equation (\ref{frd1}) can be also rewritten as the 
following relation among the Bekenstein entropy $S_B={{2\pi}\over n}Ea$, 
the Bekenstein-Hawking entropy $S_{BH}=(n-1){V\over{4Ga}}$ and the Hubble 
entropy $S_H$:
\begin{equation}
S^2_H=2S_BS_{BH}-kS^2_{BH}.
\label{entrels}
\end{equation}

We consider the moment at which the brane crosses the black hole horizon 
$a=a_H$, defined as the largest root of $h(a)=0$, i.e., 
\begin{equation}
{a^2_H\over L^2}+k-{{w_{n+1}M}\over a^{n-1}_H}=0.
\label{roothor}
\end{equation}
From Eqs. (\ref{frdeq1},\ref{roothor}), we see that
\begin{equation}
H^2={1\over L^2}\ \ \ \ \ \ \ {\rm at}\ \ \ \ \ \ a=a_H.
\label{hblhor}
\end{equation}
The entropy $S$ remains constant during the cosmological evolution, but the 
entropy density,
\begin{equation}
s={S\over V}=(n-1){a^n_H\over{4GLa^n}},
\label{entden}
\end{equation}
varies with time.  From Eqs. (\ref{hblhor},\ref{entden}), we see that the 
entropy density at $a=a_H$ is given in terms of $H$ at $a=a_H$ in the 
following form:
\begin{equation}
s=(n-1){H\over{4G}}\ \ \ \ \ \ \ {\rm at}\ \ \ \ \ \ a=a_H,
\label{entdenhor}
\end{equation}
which implies
\begin{equation}
S=S_H\ \ \ \ \ \ \ {\rm at}\ \ \ \ \ \ a=a_H.
\label{hentsat}
\end{equation}
From the temperature expression $T=h^{\prime}L/(4\pi a_H)$ at $a=a_H$ 
along with the formula $H^2=\sigma^2-h(a)/a^2$ (which follows from Eqs. 
(\ref{costim},\ref{eqbranact})), we see that the CFT temperature at $a=
a_H$ can be expressed in terms of $H$ and $\dot{H}$ in the following way:
\begin{equation}
T=-{\dot{H}\over{2\pi H}}\ \ \ \ \ \ \ {\rm at}\ \ \ \ \ \ a=a_H.
\label{tmphor}
\end{equation}
Eq. (\ref{fredcft2}) along with Eqs. (\ref{hentsat},\ref{tmphor}) implies
\begin{equation}
E_C=kE_{BH}\ \ \ \ \ \ \ {\rm at}\ \ \ \ \ \ a=a_H,
\label{enrel}
\end{equation}
where $E_C$ is the Casimir energy defined as
\begin{equation}
E_C\equiv n(E+pV-TS).
\label{casendef}
\end{equation}
So, for any values of $k$, the thermodynamic quantities of the CFT take 
the forms simply expressed in terms of the Hubble parameter and its time 
derivative when the brane crosses the black hole horizon.  

The above thermodynamic quantities of the CFT satisfy the first law of 
thermodynamics,
\begin{equation}
TdS=dE+pdV,
\label{1stlaw}
\end{equation}
which can be expressed in terms of the densities as
\begin{equation}
Tds=d\rho+n(\rho+p-Ts){{da}\over a},
\label{1stlawden}
\end{equation}
making use of $dV=nVda/a$.  If the entropy and energy are assumed to be 
purely extensive, then the combination $\rho+p-Ts$ is always zero.  For 
the conformal system under consideration, the combination is not always 
zero due to the subextensive contribution.  To find the expression for the 
combination, we express the energy density of the CFT in the following way:
\begin{equation}
\rho={{na^n_H}\over{16\pi G_{n+2}a^{n+1}}}\left({a_H\over L}+k{L\over 
a_H}\right),
\label{endenexp}
\end{equation}
and make use of the equation of state $p=\rho/n$, which is valid for CFTs.  
The resulting expression is
\begin{equation}
{n\over 2}(\rho+p-Ts)=k{\gamma\over a^2},
\label{combexp}
\end{equation}
where the Casimir quantity $\gamma$ is given by
\begin{equation}
\gamma={{n(n-1)a^{n-1}_H}\over{16\pi Ga^{n-1}}}.
\label{gamma}
\end{equation}
In other words, the Casimir energy of the CFT is given by
\begin{equation}
E_C={{kn(n-1)a^{n-1}_H{\rm Vol}(M^n)}\over{8\pi Ga}}.
\label{caseng}
\end{equation}
So, the Casimir energy is positive [negative] for $k=1$ [$k=-1$] and zero 
for $k=0$.  The entropy density (\ref{entden}) of the CFT can be expressed 
in terms of $\gamma$ and $\rho$ as
\begin{equation}
s^2=\left({{4\pi}\over n}\right)^2\gamma\left(\rho-k{\gamma\over a^2}\right).
\label{entdenexp}
\end{equation}
By making use of Eq. (\ref{entdenhor}), we can show that the entropy density 
expression (\ref{entdenexp}) at $a=a_H$ (i.e., when the brane crosses the 
black hole horizon) exactly reproduces the first Friedmann equation 
(\ref{frd1}).  Furthermore, by making use of Eqs. 
(\ref{entdenhor},\ref{tmphor}), we can show that Eq. (\ref{combexp}) 
reproduces the second Friedmann equation (\ref{frd2}) when $a=a_H$.  This 
result implies that for {\it any} values of $k$ the Friedmann equations know 
about the thermodynamic properties of the CFT.  

Since the Casimir energy (\ref{caseng}) is negative and zero respectively 
for the $k=-1,0$ cases, the Cardy-Verlinde formula proposed in Ref. 
\cite{ver1} is not valid for these cases.  We can nevertheless infer 
the modified Cardy-Verlinde formula which is valid for any $k$ from  
cosmological formulas (\ref{fredcft1},\ref{fredcft2}).  We have shown 
that $S=S_H$ and $E_C=kE_{BH}$ when $a=a_H$.  So, from the cosmological 
Cardy formula (\ref{fredcft1}) we can infer the following modified form of 
the Cardy-Verlinde formula, valid for any $k$:
\begin{equation}
S=\sqrt{{{2\pi a}\over n}S_C(2E-E_C)}.
\label{newentfrml}
\end{equation}
Since $E_C$ is zero and negative respectively for $k=0,-1$, we here choose 
to define the Casimir entropy $S_C$ first and then define $E_C$ in terms 
of $S_C$ in the following way:
\begin{equation}
S_C\equiv\left.{{2\pi}\over n}E_{BH}a\right|_{a=a_H},\ \ \ \ \ \ \ \ \ \ 
E_C\equiv{{kn}\over{2\pi a}}S_C=\left.kE_{BH}\right|_{a=a_H}{a_H\over a},
\label{defscec}
\end{equation}
where $E_{BH}$ is defined in Eq. (\ref{defs}).  
So, the Casimir entropy of the CFT is given by
\begin{equation}
S_C=(n-1){{a^{n-1}_H{\rm Vol}(M^n)}\over{4G}},
\label{casent}
\end{equation}
which is positive for any $k$, in accordance with an interpretation of 
$S_C$ as a generalization of the central charge to arbitrary spacetime 
dimensions.  Note, the definition of $E_C$ in Eq. (\ref{defscec}) is 
compatible with another definition (\ref{casendef}) of $E_C$.  This is due 
to the fact that Eq. (\ref{fredcft2}) and Eq. (\ref{casendef}) coincide 
when $a=a_H$, since $S=S_H$ and $T=T_H$ when $a=a_H$.  The modified formula 
(\ref{newentfrml}) expresses that the entropy $S$ has negative [positive] 
contribution from the Casimir effect for the $k=1$ [$k=-1$] case and no 
Casimir effect contribution for the $k=0$ case.  The modified Cardy-Verlinde 
formula (\ref{newentfrml}) can be rewritten as the following relation among 
$S$, $S_C$ and $S_B$:
\begin{equation}
S^2=2S_BS_C-kS^2_C.
\label{modcvrel}
\end{equation}
This relation has the same form as the relation (\ref{entrels}) among the 
cosmological entropy bounds, except that the roles of $S_H$ and $S_{BH}$ 
are respectively taken over by $S$ and $S_C$.  These two relations coincide 
when $a=a_H$.  

We have seen that for any values of $k$ the thermodynamic quantities of the 
CFT take the simple forms that saturate the cosmological bounds, which 
are originally conjectured \cite{ver1} for the $k=1$ case, only.  We argue 
that such conjectured cosmological bounds hold even for the $k=-1,0$ cases.  
First of all, the criterion for distinguishing between a weakly and a 
strongly self-gravitating universe becomes modified when $k\neq 1$.  If we 
choose to define the universe to be weakly [strongly] self-gravitating when 
the total energy $E$ is less [greater] than $E_{BH}$ (defined as the energy 
required to form a black hole with the size of the entire universe), then 
from the first Friedmann equation (\ref{frd1}) we see that the criterion on 
$H$ is modified to
\begin{eqnarray}
E\leq E_{BH} \ \ \ \Leftrightarrow\ \ \ S_B\leq S_{BH} 
\ \ \ \ \ \ \ \ \ &{\rm for}&\ \ \ \ \ \ \ \ Ha\leq\sqrt{2-k}
\cr
E\geq E_{BH} \ \ \ \Leftrightarrow\ \ \ S_B\geq S_{BH} 
\ \ \ \ \ \ \ \ \ &{\rm for}&\ \ \ \ \ \ \ \ Ha\geq\sqrt{2-k}.
\label{wkstrgcrt}
\end{eqnarray}
We propose that for general values of $k$ the cosmological bound on $S_C$ 
conjectured in Ref. \cite{ver1} continues to hold:
\begin{equation}
S_C\leq S_{BH}.
\label{cnjctcb}
\end{equation}
For the strongly self-gravitating case, from Eqs. 
(\ref{wkstrgcrt},\ref{cnjctcb}) we have $S_C\leq S_{BH}\leq S_B$.  From 
Eq. (\ref{modcvrel}) we see that $S$ is a monotonically increasing function 
of $S_C$ in the interval $S_C\leq S_B$ for any values of $k$.  So, $S$ 
reaches its maximum value when $S_C=S_B$ and therefore $S_C=S_{BH}$, for 
which case $S=S_H$ as can be seen from Eqs. (\ref{entrels},\ref{modcvrel}).  
The conjectured cosmological bound (\ref{cnjctcb}) on $S_C$ for the strongly 
self-gravitating case therefore implies the Hubble entropy bound for any 
values of $k$:
\begin{equation}
S\leq S_H  \ \ \ \ \ \ \ \ \ \ {\rm for}\ \ \ \ \ \ \ \ \ Ha\geq\sqrt{2-k}.
\label{entbnd}
\end{equation}
The criterion (\ref{wkstrgcrt}) on $H$ for the strongly self-gravitating 
universe along with the first Friedmann equation (\ref{frdeq1}) implies 
$a^{n-1}\leq \omega_{n+1}M/2$.  So, from the explicit expression 
(\ref{entempads}) for $T$, we infer the following cosmological bound on 
the temperature of the CFT:
\begin{equation}
T\geq T_H  \ \ \ \ \ \ \ \ \ \ {\rm for}\ \ \ \ \ \ \ \ \ Ha\geq\sqrt{2-k}.
\label{tmpbnd}
\end{equation}
These cosmological bounds (\ref{entbnd},\ref{tmpbnd}) are saturated at the 
moment when the brane crosses the black hole horizon, as shown in Eqs. 
(\ref{hentsat},\ref{tmphor}), respectively.

\end{document}